\documentclass[%
 reprint,
 amsmath,amssymb,
 aps,superscriptaddress,
]{revtex4-2}

\usepackage{amsmath}
\usepackage{xcolor}
\usepackage{graphicx}
\usepackage{dcolumn}
\usepackage{bm}
\usepackage{xcolor}

\begin{document}

\preprint{APS/123-QED}

\title{Correlating isothermal compressibility to nucleon fluctuations in the \\inner crust of neutron stars}

\author{R. Shafieepour}
\address{Department of Physics, University of Tehran, Tehran 14395-547, Iran}
\author{H. R. Moshfegh}
\address{Department of Physics, University of Tehran, Tehran 14395-547, Iran}
\address{Departamento de F\'{i}sica, Pontif\'{i}cia Universidade Cat\'{o}lica do Rio de Janeiro, Rio de Janeiro 22452-970, Brazil}
\author{J. Piekarewicz}
\address{Department of Physics, Florida State University, Tallahassee, Florida 32306-4350, USA}

\date{\today}

\begin{abstract}
The question of how and which physical observables or thermodynamic parameters can best predict the onset of a possible phase transition in the inner crust of neutron stars remains largely unresolved. Using semiclassical Monte Carlo simulations, we investigate the isothermal compressibility and density fluctuations in a region of relevance to the dynamics of the inner crust. We show that the isothermal compressibility serves as a robust observable to characterize the transition from the non-uniform crust to the uniform core for proton fractions over 0.2. Moreover, we show explicitly how the two-component isothermal compressibility, computed using the Kirkwood-Buff theory, is directly connected to the fluctuations in the number density, recorded in the grand canonical ensemble by monitoring the number of particles in a small volume located at the center of the simulation box. That is, we compute mean-square particle fluctuations and compare them against the isothermal compressibility for different proton fractions. Although our results show that the mean-square particle fluctuations are proportional to the isothermal compressibility, the lack of a perfect correlation is attributed to the relatively small number of particles included in the simulations. The non-unity slope observed in the dimensionless isothermal compressibility–total nucleon fluctuation variance relationship suggests that the inner crust of neutron stars is composed of anisotropic and inhomogeneous matter.
\end{abstract}

\maketitle
\section{INTRODUCTION}

Shortly after being discovered by Bell and Hewish in 1967\,\cite{1}, neutron stars have been identified as unique laboratories for the study of 
hadronic matter under extreme conditions. Given their enormous density range, it is well-accepted that neutron stars are stratified into a low-density
non-uniform crust and a homogeneous core, both embedded in a uniform neutralizing leptonic Fermi gas\,\cite{2,3}. Given that the baryon density 
varies by more than five orders of magnitude from the inner core to the outer crust\,\cite{4}, it is anticipated that the transition from a Coulomb crystal
of neutron-rich nuclei at low densities to the homogeneous core is accompanied by a dramatic change in the topology of the nuclear clusters, the
so-called pasta phases\,\cite{5,6}. In addition, as the transition density to uniform matter is predicted to occur between a third and a half of nuclear
matter saturation density, investigations of the crustal region could inform how a neutron-rich skin emerges in neutron-rich nuclei such as ${}^{48}$Ca
and ${}^{208}$Pb nucleus\,\cite{7}. 

It was back in 2004 when Horowitz \textit{et al.} kindled the first sparks of the possible impact of the pasta phases on neutrino transport in supernovae 
and proto-neutron stars\,\cite{8}. Traditionally, both Monte-Carlo (MC) and Molecular-Dynamic (MD) simulations have been used to simulate the complex 
dynamics for a system displaying Coulomb frustration---a universal phenomenon that emerges from the competition between attractive short-range 
interactions and the long-range Coulomb repulsion. Numerical simulations of this kind have been carried out as a function of baryon density $\rho$, 
proton fraction $y_{p}$, and temperature $T$, for a range of values that span a region in the inner crust where the pasta phases are expected to 
emerge\,\cite{9,10,11,12,13,14}. The significant impact of the pasta phases on the dynamics of the inner crust has been invoked to explain the 
lack of x-ray-emitting isolated pulsars with long spin periods, pulsar glitches, the elasticity of the inner crust, the possibility of enhanced neutrino 
cooling via the direct Urca process, the delay in the arrival of the neutrino signal, and on magnetic field decay. For instance, Pons \textit{et al.} unveiled 
that a highly resistive layer in the inner crust limits the spin period to a maximum value of about 10-20 seconds\,\cite{15}. In turn, Piekarewicz \textit{et al.} 
found that uncertainties in the equation of state of neutron-rich matter are large enough to accommodate theoretical models that predict large fractal 
crustal moments of inertia that are essential to explain pulsar glitches\,\cite{16}. Moreover, using MD simulations with a large number of particles, 
Caplan \textit{et al.} studied the breaking mechanism of idealized nuclear pasta plates by applying tensile and shear strains for specific values of
the density and proton fraction of $\rho\!=$ 0.05 fm$^{-3}$ and $y_{p}\!=0.4$, respectively.  Their results suggest that nuclear pasta may be the 
strongest known material, with a shear modulus larger than 0.1 and breaking strain of 10$^{30}$ ergs cm$^{-3}$\,\cite{17}. Finally, by modeling a 
variety of pasta phases with different topologies, Lin \textit{et al.} demonstrated that the neutrino luminosity from the direct Urca process in the inner 
crust can be 3-4 orders of magnitude larger than that from the modified Urca process in the stellar core\,\cite{18}. In a subsequent work, it was 
concluded that the scattering of neutrinos from the complex pasta structures may slow their diffusion, thereby increasing the late-times neutrino 
signal from the collapse of the stellar core\,\cite{19}.

Further, significant resources have been devoted to identifying physical observables sensitive to the crust-core transition, a task that has proven to be
highly challenging. For instance, and departing from the MC/MD paradigm, Burrello \textit{et al.} presented results obtained in a quasiparticle 
mean-field Hartree-Fock-Bogoliubov (HFB) theory in the Wigner-Seitz approximation\,\cite{20}. Their results revealed a peak in the heat capacity 
at the critical temperature at which the nuclei melt into a gas of free particles and resonances in the inner crust. Although properly incorporating
quantum effects, mean-field approaches miss some of the complex clustering correlations that are properly captured in the semi-classical approach. 
Instead, using a quantum MD simulation in which the single-nucleon wave functions are represented by Gaussian wave packets, Nandi and 
Schramm reported that various transport properties in the inner crust---such as the electrical and thermal conductivity as well as the shear 
viscosity---are of the same order of magnitude as those found without the pasta phase\,\cite{21}. These results call into question the reliability 
of such transport properties in identifying exotic behavior associated with the existence of the pasta phases. Recently, we reported using an MC
simulation that the isothermal compressibility computed in the framework of the Kirkwood-Buff theory, reaches a maximum when isolated 
non-symmetric clusters are formed in an extremely dilute neutron gas at $\rho\!=$ 0.008 and 0.01 fm$^{-3}$ and for $y_{p}\!=\!0.4$ and 0.2, 
respectively\,\cite{22}. Such behavior would be reminiscent of a critical point (albeit not explicitly) associated with a maximum in the density 
fluctuations, a phenomenon that has been extensively investigated in the phase diagram of conventional liquids such as water. However, our 
previous simulation results are limited by the numerical challenge of generating numerous configurations covering the vast thermodynamic 
conditions that exist within the inner crust. 

In the present work, we perform MC simulations for a wide range of temperatures, baryon densities, and proton fractions. Based on these 
results, we propose the two-component isothermal compressibility as a robust thermodynamic observable for identifying the onset of the 
phase transition. In turn, we correlate the isothermal compressibility in the critical region to the variance in the nucleon number. Computing
density fluctuations is implemented here by selecting a ``small" portion of the simulation box as the system of interest and letting the ``large" 
portion of the box acts as the particle reservoir. To our knowledge, no previous work has attempted to connect the isothermal compressibility
to the density fluctuations in such a manner.  

The paper is organized as follows. Sec. II describes the temperature-dependent semi-classical MC simulation, the theoretical formalism used 
to calculate the isothermal compressibility, and our proposed method to record the variance in the number density. Sec. III is devoted to a
discussion of our results for the evolution of the isothermal compressibility and its connection to the variance in the density. Moreover,
in this section, we illustrate how these observables serve to identify the onset of the phase transition. Finally, we offer our conclusions in 
Sec. IV.

\section{FORMALISM}
\label{sec:Formalism} 
In this section, we describe the computational and theoretical formalism that will be used to simulate nuclear matter under the thermodynamic
conditions relevant to the inner crust of neutron stars. We commence by providing a concise overview of the Monte Carlo simulation. The rest 
of this section is dedicated to the computation of various observables associated with neutron stars.

\subsection{Semiclassical Monte Carlo simulation}
The MC simulation employed in this study follows closely the methodology implemented in our recent publication\,\cite{22}. Yet, for the sake of 
clarity, we present a brief overview of the key points. For all the simulations we adopt a fixed number of $A\!=\!5,000$ nucleons. Hence, to account for baryon densities spanning the 0.08--0.005 fm$^{-3}$ range, we must vary the box length $L$ from 39.69 to 100 fm, respectively. In 
turn, the proton fraction were set at three fixed values, namely, $y\!=Z/A\!=$ 0.1, 0.2, and 0.4, where $Z$ represents the number of protons, which
in all cases equals the number of neutralizing electrons that are treated as a noninteracting Fermi gas. 

The total potential energy of the $A$-body system is given in terms of the following two contributions:
\begin{equation}
V({\bf r}_{1},\ldots,{\bf r}_{A})\!=\!V_{\rm N}({\bf r}_{1},\ldots,{\bf r}_{A})+V_{\rm C}({\bf r}_{1},\ldots,{\bf r}_{A}),
\label{eq:E1}
\end{equation}
where $V_{\rm N}$ and $V_{\rm C}$ stand for the short-range nuclear and long-range Coulomb potentials, respectively. The vector $\bf r_{i}$ denotes 
the position of the nucleon labeled with the index $i$. The functional form of the underlying nucleon-nucleon potential, the implementation of the Ewald 
summation for a periodic box, and details of the Metropolis algorithm can all be found in earlier reports\,\cite{11, 23}. The use of periodic boundary 
conditions is used to mitigate finite-size effects associated with the limited size of the simulation box.

To start the MC simulation, we select an initial temperature of $T\!=\!2$\,MeV and with all $A\!=\!5,000$ nucleons distributed randomly throughout the 
box. Once the initial thermalization phase has been completed, the system is cooled gradually to a final temperature of $T\!=\!0.5$\,MeV using a cooling 
schedule of 0.1 MeV per 4,000 sweeps. We note that each sweep consists of $A$ individual MC steps so that on average each nucleon is tagged one
time. Once the final temperature is reached, an additional 50,000 sweeps are performed to ensure full thermalization of the system. The simulation
concludes with 5,000 additional sweeps to accumulate enough statistics on the relevant physical observables. A similar process is followed in reverse
as we investigate the dynamics at higher temperatures. In this case, starting at $T\!=\!0.5$\,MeV, the temperature increases in steps of 0.25 MeV until 
the target temperature is reached. At each temperature stage, we use 50,000 thermalization sweeps and finish with 5,000 additional sweeps to collect
statistics at the target temperature.

\subsection{Isothermal compressibility}
As in our recent work\,\cite{22}, we employ a formalism developed by Kirkwood and Buff more than 70 years ago to compute the isothermal 
compressibility $\kappa_T$ of a system consisting of an arbitrarily large number of constituents\,\cite{23}. Concisely, Kirkwood and Buff 
developed a general method based on the theory of composition fluctuations in the grand canonical ensemble to compute the isothermal 
compressibility for an m-component system\,\cite{23}. In the particular case of a 2-component system as the one considered in this work, 
the isothermal compressibility is given by the following expression\,\cite{22}:
\begin{equation}
\rho\,T\kappa_T\!=\frac{1\!+\!\rho_{n}G_{nn}\!+\!\rho_{p}G_{pp}\!+\!\rho_{n}\rho_{p}(G_{nn} G_{pp}\!-\!G_{np}^2)}
           {\left[1+\displaystyle{\left(\frac{\rho_{n}\rho_{p}}{\rho}\right)}\big(G_{nn}\!+\!G_{pp}\!-\!2G_{np}\big)\right]},
\label{eq:E2}
\end{equation}
where $G_{ij}$ is the angle-averaged integral of the pair correlation function for species $i$ and $j$, namely,
\begin{equation}
 G_{ij}\!=\displaystyle \int \limits_{0}^{\infty} \big[g_{ij}(r)-1\big]4\pi r^2 dr,
 \label{eq:E3}
\end{equation}
and where $g_{ij}(r)$ is the radial distribution (or  pair correlation) function given by
\begin{equation}
 g_{ij}(r)\!=\!\frac{L^3}{4\pi r^2 N_{i} N_{j}}
 \left\langle\sum_{i}\sum_{j\neq i}\delta\Big(r-\mid {\bf r}_i\!-\!{\bf r}_j \mid\Big)\right\rangle.
\label{eq:E4}
\end{equation}
Here the brackets indicate an ensemble average over all MC configurations and $g_{\ij}(r)$ is normalized to unity as $r$ approaches $L/2$. 
The pair correlation function is both an insightful and important observable. Indeed, for pairwise interactions, as we assume here, fundamental 
thermodynamic observables, such as the internal energy, pressure---and, of course, the isothermal compressibility---may be readily obtained 
by performing suitable integrals involving the pair correlation function. Moreover, given that in MC simulations one keeps track of the positions 
of all the particles, $g_{\ij}(r)$ is an observable that is relatively simple to compute. 


\begin{figure}[h]
	\begin{center}
		\includegraphics[trim={0 0 0 0},scale=1,width=8.6cm]{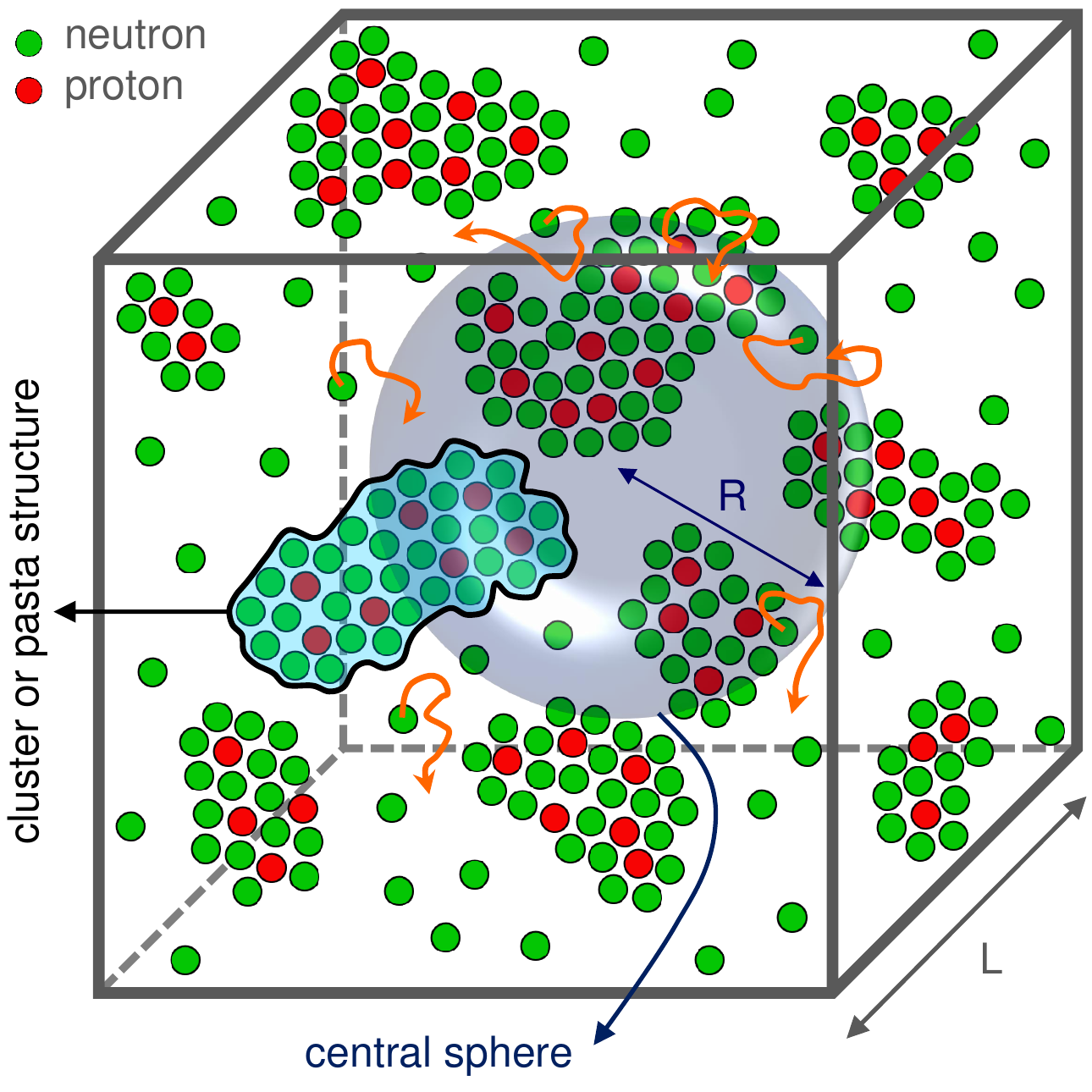}
		\caption{Schematic illustration of a configuration confined to a simulation box of length $L$. For large systems,  
		              one could in principle implement a grand canonical formulation by adopting the central sphere of radius $R$ as 
		              the system of interest and letting the rest of the box act as the particle reservoir. The orange arrows show the 
		              possible trajectories of some nucleons during the MC simulation.}
		\label{Figure_1}
	\end{center}
\end{figure}

\subsection{Particle number fluctuations}

Given that to this date most MC and MD simulations have been carried out in the canonical ensemble where the number of particles, 
the temperature, and the volume are the control thermodynamic variables, how does one compute fluctuations in the number of particles 
in such a scenario? To do so we introduce a novel scheme which, as shown in Fig.\;\ref{Figure_1}, consists of dividing the original 
simulation box into two regions: a central sphere of radius $R\!=\!0.36\,L$ containing approximately 20\% of the original volume and a 
particle reservoir containing the remaining 80\% and responsible for enforcing chemical equilibrium. As such, one monitors the number 
of protons $N_{p}(s)$, neutrons $N_{n}(s)$, and the total number of nucleons $N_{t}(s)\!=\!N_{p}(s)\!+\!N_{n}(s)$ in the smaller volume 
as a function of the Monte Carlo step $s$, and then compute suitable averages over the total number of sweeps $S\!=\!5,000$. The 
fluctuations in the particle number of a given species ($\mu\!=\!p,n,t$) are then given by
\begin{equation}
 \sigma_{\mu}^{2}\!=\!\frac{1}{S}\sum_{s=1}^{S}\Big(N_{\mu}(s)\!-\!\langle N_{\mu} \rangle\Big)^{2} 
                        \!=\!\langle N_{\mu}^{2} \rangle\!-\!\langle N_{\mu} \rangle^{2}.
 \label{eq:E5}
\end{equation}

We conclude this section by connecting three seemingly distinct physical observables encoding: (\textit{i}) the structural properties of 
the system $S(q)$, (\textit{ii}) the underlying equation of state $\kappa_{T}$, and (\textit{iii}) the statistical fluctuations in the number 
of particles $\langle N^{2}\rangle\!-\!\langle N\rangle^{2}$. That is,
\begin{equation}
 S(q\!=\!0) = \rho\hspace{1pt}T\kappa_{T} = \frac{\langle N^{2}\rangle\!-\!\langle N\rangle^{2}}{\langle N\rangle},
 \label{eq:E6}
\end{equation}
where $\kappa_{T}$ is obtained from the pair-correlation function as indicated in Eq.(\ref{eq:E2}). Consistency among all these quantities 
provides a robust and non-trivial test of the entire formalism.

\section{RESULTS AND DISCUSSIONS}

\begin{figure*}[hbt]
	\begin{center}
		\includegraphics[trim={0 0 0 0},scale=1,width=17.6cm]{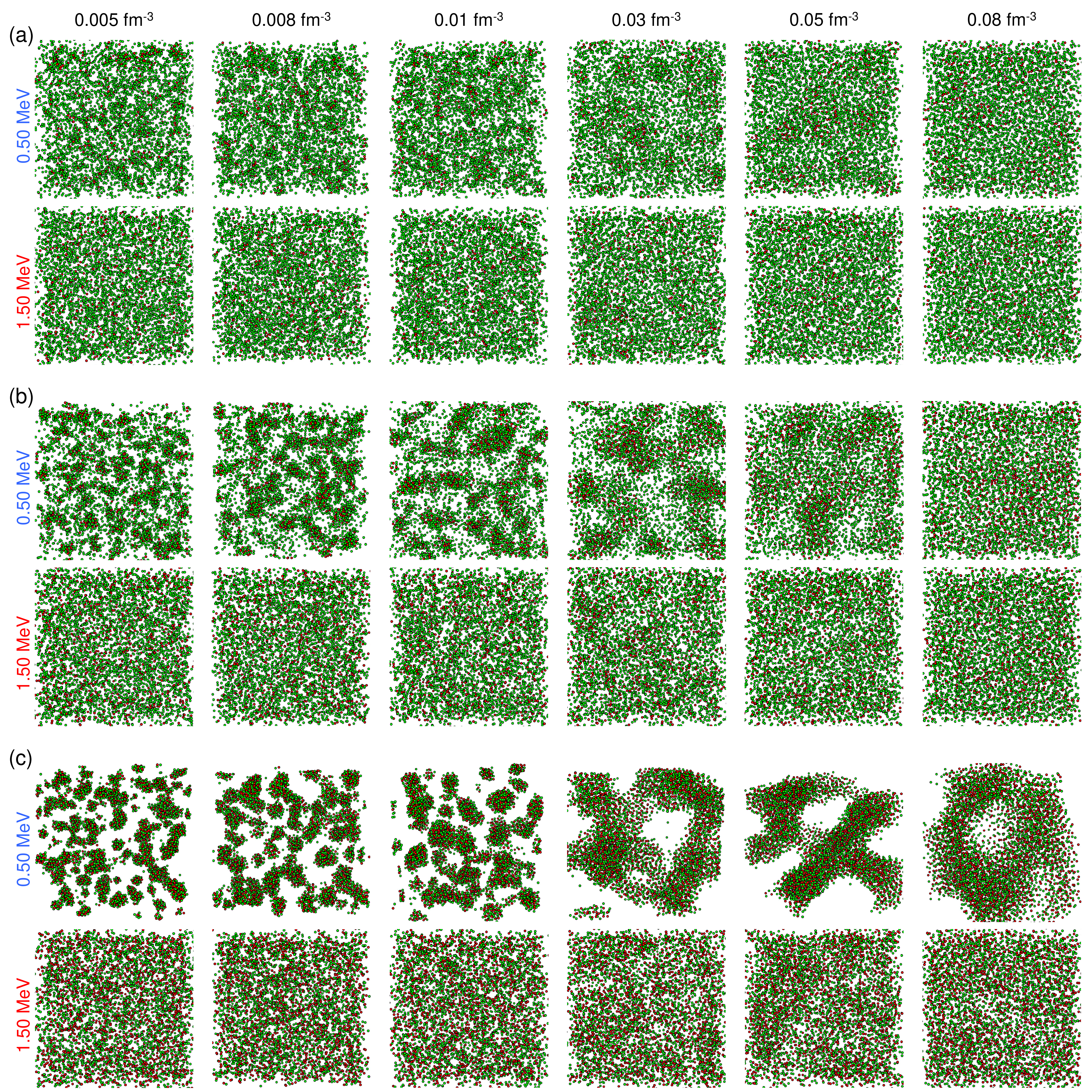}
		\caption{Snapshots of thermalized nucleon configurations for a variety of baryon densities, at proton fractions of (a) $0.1$, (b) $0.2$, 
		and (c) $0.4$, and a temperature of $T\!=\!0.5$ and $1.5$\,MeV. Red and green solid circles depict protons and neutrons, respectively. 
		For space limitations, we do not show nucleon configurations at the intermediate temperature range of $T\!=0.75–1.25$.} 
         \label{Figure_2}
	\end{center}
\end{figure*}

In this section, we present an analysis of our MC simulations to identify the onset of the phase transition from non-uniform 
(``clustered") matter in the inner crust to uniform neutron-rich matter in the stellar core. 

\subsection{Thermalized configurations}
\begin{figure*}[hbt]
	\begin{center}
		\includegraphics[trim={0 0 0 0},scale=1,width=17.6cm]{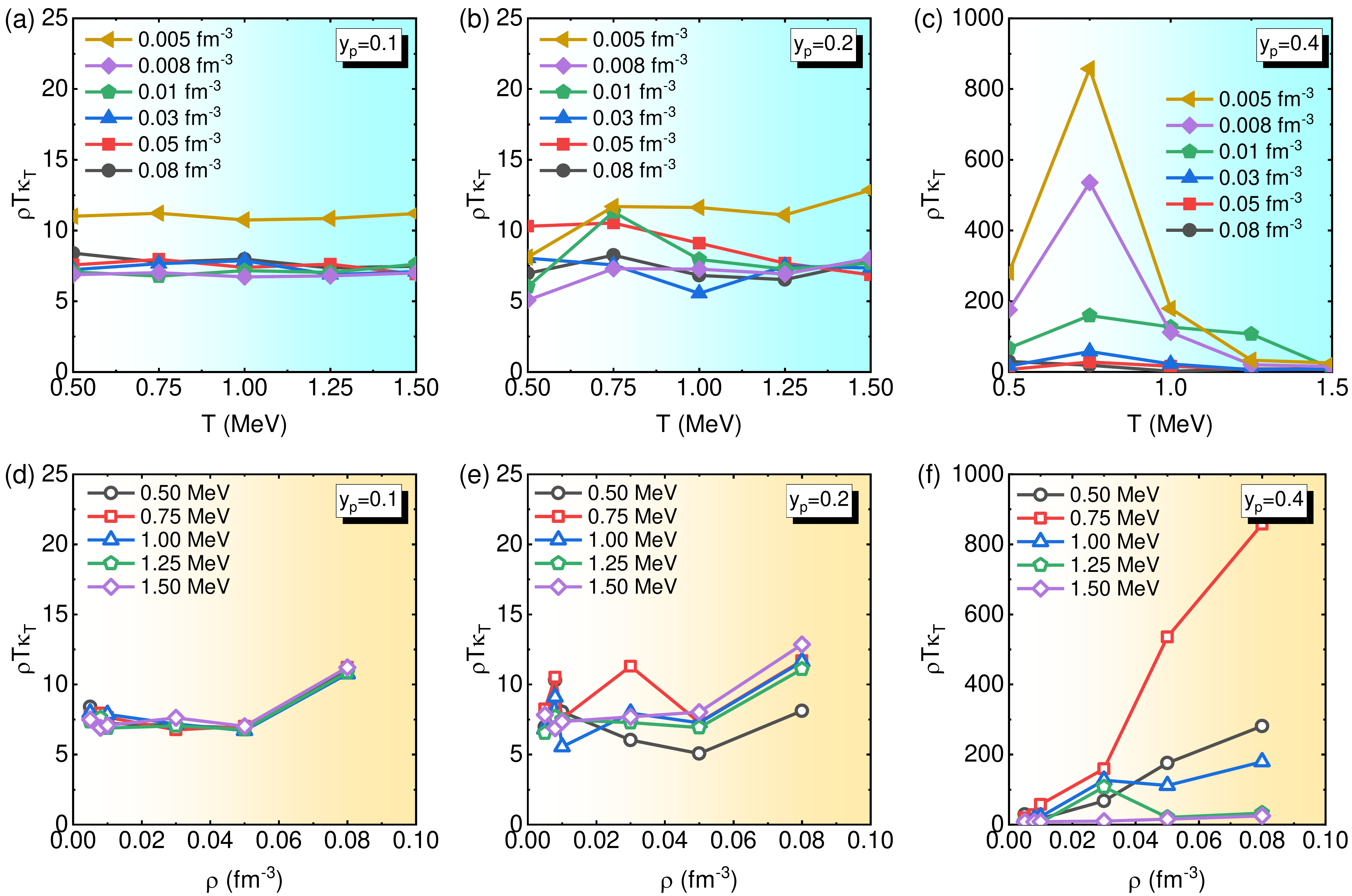}
		\caption{$\rho T \kappa_{T}$ versus (a)–(c) $T$ and (d)–(f) $\rho$ for different proton fractions of $0.1$, $0.2$, and $0.4$, respectively. The various lines are added to guide the eye.} \label{Figure_3}
	\end{center}
\end{figure*}
We start by displaying in Fig.\;\ref{Figure_2} snapshots of various thermalized nucleon configurations under different thermodynamic conditions.
The various configurations were obtained at two temperatures  $T\!=\!0.5, 1.5$ MeV, and for the lower temperature they are classified into four 
categories as follows:

\begin{enumerate}

\item[(\textit{i})]  $y_{p}\!\leq\!0.2$ for $\rho\!\leq\!0.01$ fm$^{-3}$ and $y_{p}\!=0.1$ for $\rho\geq\!0.03$ fm$^{-3}$. 
 In this case, the thermodynamic conditions favor the formation of nonsymmetric isolated clusters immersed in a uniform neutron vapor. This 
 situation is reminiscent of the distillation effect, corresponding to the formation of high-density clusters of nearly symmetric matter immersed
 in a background consisting of a neutron gas and possibly, also containing a very small fraction of protons\,\cite{24}. Further, we note that the size 
 of the clusters increases both with density and proton fraction. A larger proton fraction binds more neutrons into the cluster due to the nuclear 
 attraction between neutrons and protons.

\item[(\textit{ii})] $y_{p}\!=0.2$ for $\rho\geq\!0.03$ fm$^{-3}$. 
The conditions are such that the moderate density and relatively large proton fraction favor the formation of elongated and nonsymmetric 
clusters that are starting to touch. As in isolated nuclei with a large proton content, the long-range Coulomb interaction induces the deformation 
of the cluster. If many such clusters coexist, then exotic shapes with different topologies start to emerge. One can also see that at this density
some of the large clusters remain well-separated, suggesting that under these conditions the system contains a mixture of isolated deformed, pasta phases, and a neutron vapor.

\item[(\textit{iii})] $y_{p}\!=\!0.4$ for $\rho\leq\!0.01$ fm$^{-3}$. 
For such a dilute system, the clusters are well separated. Moreover, because of the large proton fraction, the neutron vapor largely disappears 
and nearly all neutrons are absorbed into clusters that take the shape of well-defined nuclei. In turn, because at large separations only the
Coulomb interaction between clusters remains effective, the system organizes itself into a Coulomb crystal. These results agree with 
our previous work where we have shown that more protons can bind a larger fraction of neutrons\,\cite{22}.

\item [(\textit{iv})] $y_{p}\!=0.4\!$ for $\rho\geq\!0.03$ fm$^{-3}$. 
These conditions are optimal for the development of the pasta phases, as the large proton fraction induces the deformation of the clusters
whereas the high density forces them to overlap and merge---ultimately yielding exotic shapes of various topologies. Again, these results 
are well-matched with our previous report\,\cite{22}.

\end{enumerate}

As the temperature increases to $T\!=1.5$ MeV, the loosely bound neutrons separate from the clusters, resulting in tightly bound nuclei 
immersed in a uniform neutron vapor at all proton fractions. We observe the incomplete liquefaction (or melting) of the clusters and pasta 
phases, suggesting that only a portion of the nuclear material has turned into a vapor, while the rest remains in its original state. Therefore, 
one can still see the presence of nuclear clusters and pasta phases even at this increased temperature. Given the drastic change in the
structure of the system with increasing temperature, we anticipate a dramatic change in the physical and thermodynamic properties of 
the system, indicating the onset of a phase transition. In the following sections, we will explore the impact of the temperature on both the 
isothermal compressibility and the fluctuations in the number of particles.

\subsection{Isothermal compressibility}
\label{Sec:IIIB}
The main observable discussed in this section is the isothermal compressibility:
\begin{equation}
  \kappa_{T} = -\frac{1}{V}\left(\frac{\partial V}{\partial P}\right)_{\!T}
             	     = \frac{1}{\rho}\left(\frac{\partial \rho}{\partial P}\right)_{\!T},
 \label{eq:E7}
\end{equation}
where $P$ is the pressure and $V$ is the volume. In particular, we study the isothermal compressibility given by the Kirkwood-Buff formula 
displayed in Eq.(\ref{eq:E2}).  In Fig.\;\ref{Figure_3} we show results for the product $\rho T \kappa_{T}$ as a function of $T$ and $\rho$ for 
proton fractions of $y_{p}\!=0.1$, $0.2$, and $0.4$. Note that for a classical ideal gas the product $\rho T \kappa_{T}$ is equal to one, so 
by plotting it in this manner one isolates the non-trivial thermodynamic behavior of the pair correlation function; see Eq.(\ref{eq:E2}). Motivated 
by some of the trends observed above, we classify this data into two categories: $y_{p}\!\leq\!0.2$ and $y_{p}\!=\!0.4$.

\begin{enumerate}
\item[(\textit{i})] $y_{p}\!\leq\!0.2$. The results of the simulations indicate that although significantly different than the ideal 
gas limit, $\rho T \kappa_{T}$ displays a mild dependence on both temperature and density. At these densities, all protons are contained
in relatively small clusters that, according to the liquid drop formula, are largely incompressible. In addition, neutrons in the vapor remain
separated because of the strong repulsion at short distances. Qualititaively, the neutrons in the vapor resemble a van der Waals (vdW) gas 
interacting via an excluded-volume (repulsive) term and an attractive dipole-dipole interaction that falls as a power law at large distances. 
Quantitatively, however, there are important differences in the nuclear case, as neutrons repel at short distances but the attraction at
intermediate distances falls exponentially.  Nevertheless, the observed behavior demonstrates that the isothermal compressibility is 
an inverse function of baryon density and temperature. For instance, as the density increases, the system evolves into larger and elongated 
non-symmetric clusters that are difficult to compress. In turn, an increase in temperature leads to a larger fraction of neutrons in the vapor
resulting in a higher background pressure. Such a reduction in the isothermal compressibility with increasing temperature has been previously 
documented in various systems, including a hot and dense hadron gas and supercooled water\,\cite{25, 26}.
\begin{figure*}[hbt]
	\begin{center}
		\includegraphics[trim={0 0 0 0},scale=1,width=17.6cm]{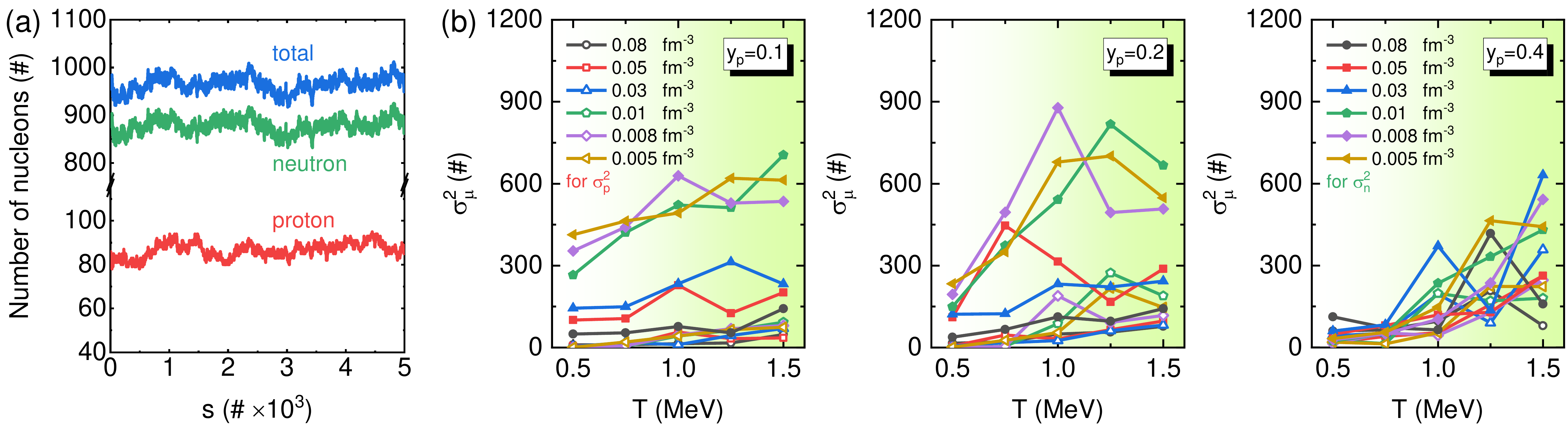}
		\caption{(a) An example for the number of nucleons inside the central sphere as a function of $s$ for 
		$\rho\!=0.03$ fm$^{-3}$ and $y_{p}\!=0.1$ at $T\!=1$ MeV. (b) Variance in the neutron ($\sigma_{n}^{2}$, 
		filled symbols) and proton ($\sigma_{p}^{2}$, empty symbols) number versus temperature for different 
		proton fractions of $0.1$ (left panel), $0.2$ (middle panel), and $0.4$ (right panel). The various lines are 
		added to guide the eye.} \label{Figure_4}
	\end{center}
\end{figure*}
\item[(\textit{ii})] $y_{p}\!=0.4$. The isothermal compressibility of such a nearly isospin-symmetric system displays different behavior. First, 
however, we indicate how $\rho T \kappa_{T}$ displays a similar trend as that observed at low proton fraction for $\rho\!=0.08$ fm$^{-3}$ 
and $T\!=1.5$ MeV;  see Figs.\;\ref{Figure_3}c-f). The low isothermal compressibility of these two curves suggests that the Coulomb-driven
pasta phases at $\rho\!=0.08$ fm$^{-3}$ and the pressure support of the neutron vapor at $T\!=1.5$ MeV keep the system relatively 
incompressible (see Fig.\;\ref{Figure_2}). However, at the lowest density of $\rho\!=0.005$ fm$^{-3}$, Fig.\;\ref{Figure_3}(c) displays large
fluctuations with temperature. Given that at such a large proton fraction many of the neutrons in the vapor migrate to the clusters, the
reduction in the pressure support from the free neutrons results in a higher compressibility. Indeed, in the density interval 
$0.008\,{\rm fm}^{-3}\!\leq\!\rho\!\leq\!0.05\,{\rm fm}^{-3}$, Fig.\;\ref{Figure_3}(c) indicates the presence of a maximum value at $T\!=0.75$ MeV 
that may be reminiscent of a critical point where the phase boundary between free neutrons and clusters/pasta structures vanishes. It is worth 
noting that the limited data points on temperature and the limited box sizes are the main reasons why we do not approach the exact critical
temperature $T_{c}$ where $\kappa_{T}$ would diverge in the thermodynamic limit. Nevertheless, we suggest that the isothermal compressibility 
diverges at $T_{c}$ according to,
\begin{equation}
\kappa_{T}\!=\left|\frac{T}{T_{c}}-1\right|^{\!-\gamma},
\label{eq:E8}
\end{equation}
where the ``critical" exponent $\gamma$ is a positive constant, as reported for terrestrial materials \cite{27,28}. \end{enumerate}

In closing this section, it is essential to highlight the profound interplay between different topological structures and thermodynamic parameters. 
Our forthcoming research endeavors to simulate and analyze these intricate relationships, employing a substantially larger number of particles 
to explore and generate diverse pasta phases such as waffles, spaghetti, gnocchi, and lasagna\,\cite{29,30}. Following this, we intend to apply 
the same methodology to investigate the temperature-dependent two-component isothermal compressibility, aiming to gain deeper insights into 
the inner crust.

\subsection{Nucleon Fluctuations}
\label{Sec:IIIC}

So far, insights into isothermal compressibility have been developed through its connection to the pair correlation function, a fundamental
quantity that is relatively easy to obtain from MC simulations. However, much more difficult to simulate in the canonical ensemble employed 
here are density fluctuations which, as shown in Eq.(\ref{eq:E6}), are also encoded in the isothermal compressibility. To mitigate this problem, 
we have divided the simulation volume into a ``small" central sphere of volume $V_{\rm s}\!=\!4\pi R^{3}/3$ and a particle reservoir with a 
``large" volume $V\!-\!V_{s}$. In this manner, nucleon fluctuations may be quantified by monitoring the number of protons $N_{p}(s)$ and 
neutrons $N_{n}(s)$ contained within the central sphere as a function of MC-step $s$. It is important to acknowledge that while the framework 
proposed here holds promise for much larger systems, the simulations conducted here using a modest number of $A\!=\!5,000$ particles, may 
not adequately capture the fluctuations in the number of particles.

The Monte-Carlo history of the number of protons $N_{p}(s)$, neutrons $N_{n}(s)$, and total number of nucleons $N_{t}(s)$ inside the 
central sphere as a function of MC step $s$ is displayed in Fig.\;\ref{Figure_4}(a) for the following thermodynamic conditions: 
$\rho\!=\!0.03\,{\rm fm}^{-3}$, $y_{p}\!=\!0.1$,  and $T\!=\!1\,{\rm MeV}$. The observed nucleon fluctuations are attributed to free neutrons 
entering and leaving the sphere, bound neutrons/protons jiggling inside the clusters/pasta structures, free neutron absorption, bound neutron 
desorption, and clusters/pasta structures displacements. Notably, the contribution from the latter is small given that the recorded pair correlation 
functions remain largely constant---especially at longer distances---indicating that the long-range order does not change appreciably once 
the system is well-thermalized. It's worth mentioning that the observed trends highlighted here are consistent with other configurations 
investigated under different thermodynamic conditions. 

Displayed in Fig.\;\ref{Figure_4}(b) as a function of temperature for different proton fractions are the fluctuations in the number of particles
$\sigma^{2}_{\mu}$ (with $\mu\!=\!n,p$) inside the spherical simulation volume. For a density of $\rho\!=\!0.03\,{\rm fm}^{-3}$, these
fluctuations are encoded in the behavior displayed in Fig.\;\ref{Figure_4}(a). The results illustrate some general trends as well as some specific 
behavior depending on the proton fraction. Some of the observed general trends are the enhancement of $\sigma^{2}_{\mu}$ as a function of 
$T$ as well as a systematic enhancement of $\sigma_{n}^{2}$ relative to $\sigma_{p}^{2}$. Such an enhancement is due to the limited proton 
mobility, as most of the protons are confined within clusters whereas neutrons can also be found in the surrounding vapor. To further elucidate 
the behavior of the particle fluctuations, we divide the forthcoming discussion into two regions: 
(\textit{i}) $y_{p}\!\leq\!0.2$ and (\textit{ii}) $y_{p}\!=\!0.4$.

\begin{enumerate}
\item[(\textit{i})] $y_{p}\!\leq\!0.2$. The collected data points in this region exhibit a jump in $\sigma_{n}^{2}$ for densities above
$\rho\!=\!0.03\,{\rm fm}^{-3}$. Given the moderate densities and low proton fractions, such behavior can be associated with the jiggling 
of the bound neutrons within the clusters or the continuous absorption into or desorption from the clusters. Notably, as seen in 
Fig.\,\ref{Figure_4}(b), the free neutrons do not contribute significantly to the jump As the proton fraction increases from $\rho\!=\!0.01$ 
to $\rho\!=\!0.02\,{\rm fm}^{-3}$, the fraction of neutrons bound to clusters also increases. Given that neutrons under these conditions
are weakly bound, neutron absorption and desorption are strongly enhanced. 

\item[(\textit{i})] $y_{p}\!=0.4$: The data points in this region significantly decline in $\sigma_{n}^{2}$ and a moderate increase in 
$\sigma_{p}^{2}$ due to neutron localization and proton delocalization, respectively. In a proton-rich environment, proton 
delocalization may arise from the clear emergence of Coulomb frustration that favors the formation of exotic pasta phases where
the Coulomb repulsion plays a predominant role that results in a weak binding of the protons to the clusters. These quantitative
results encapsulate the qualitative picture displayed in Fig.\;\ref{Figure_2} for a representative set of MC configurations.
\end{enumerate}

\subsection{Correlating density fluctuations to the isothermal compressibility}

In this section, we aim to investigate how the particle fluctuations discussed in the previous section correlate to the behavior displayed by the isothermal compressibility computed from the Kirkwood-Buff theory. In Sec.\ref{sec:Formalism} we indicated how the isothermal compressibility defined in Eq.(\ref{eq:E7}) in terms of thermodynamic variables may be determined by computing either the pair correlation function as in Eq.(\ref{eq:E2}) or the mean square density fluctuations as in Eq.(\ref{eq:E6})\,\cite{31}. While the computation of pair correlation functions using MC or MD simulations has become routine,  we have outlined here a procedure by which one may compute density fluctuations---even within the canonical ensemble. This procedure involves monitoring the fluctuations in the number of particles entering and exiting the confined spherical volume depicted in 
Fig.\;\ref{Figure_1}, which constitutes a small fraction of the total volume of the system. The fluctuations in the number of neutrons $\sigma_{n}^{2}$ and protons $\sigma_{p}^{2}$ were the primary focus of the preceding section.

\begin{figure}[t]
	\begin{center}
		\includegraphics[trim={0 0 0 0},scale=1,width=8.6cm]{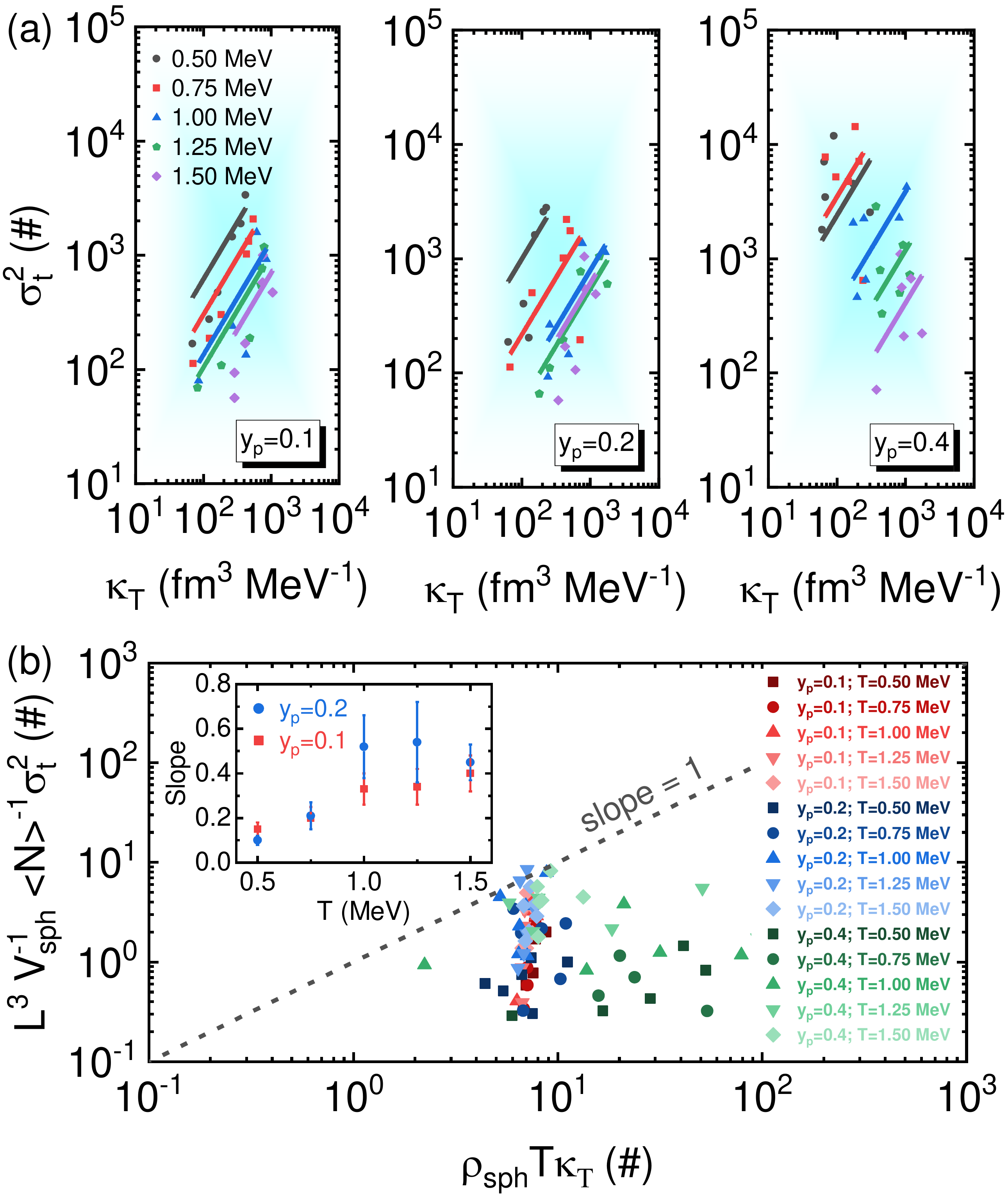}
		\caption{(a) Mean-square nucleon fluctuations versus isothermal compressibility for various temperatures at a proton fraction of $0.1$, $0.2$, and $0.4$, respectively; solid lines are to guide the eye. (b) $L^{3}V_{sph}^{-1}<N>^{-1}\sigma_{t}^{2}$ versus $\rho_{sph} T\kappa_{T}$ for all studied conditions. The gray dashed line stands for a slope of unity. Inset shows the slope of the fitted line for $y_{p}=0.1$ and $0.2$ as a function of simulation temperature.}
		\label{Figure_5}
	\end{center}
\end{figure}

We now proceed to correlate the behavior of the isothermal compressibility $\kappa_{T}$ computed within the Kirkwood-Buff 
framework with the fluctuations in the total particle number $\sigma_{t}^{2}$. It is important to mention that such a correlation exists at each temperature, baryon density and proton fraction. Fig.\;\ref{Figure_5}a represents the variation of the total number of nucleons as a function of isothermal compressibility for different proton fractions at various temperatures. We notice a correlation between these two observables, indicating the presence of a linear relationship (a line with the assumed $y$-intercept of zero) between them. It is important to note that the plotted data exhibits a significant amount of scattering and that the correlation coefficients are relatively low, likely attributable to the limited number of nucleons used in the simulation. Nevertheless, one can infer a general trend between these two quantities,
\begin{equation}
	\kappa_{T}\propto \sigma_{t}^{2},
\label{eq:E9}
\end{equation}
implying that the total particle fluctuation in the confined volume is correlated with the isothermal compressibility obtained by applying the Kirkwood-Buff theory encoded in  Eq.(\ref{eq:E6}). It is important to emphasize that the determination of the precise relationship between these two quantities, potentially in the form of $\kappa_{T} = \sigma_{t}^{2} \times F(\rho, y_{p}, T)$, where $F(\rho, y_{p}, T)$ represents a function that predicts their exact correlation at each temperature, baryon density, and proton fraction requires at least an order of magnitude increase in the number of particles. Only then one can investigate, not only the approach to the thermodynamic limit but also record the fluctuations in the number of particles within several restricted volumes, thereby enhancing the robustness and reliability of the statistical analysis.  
Nevertheless, one should also consider that the simple relation between the thermal compressibility and the fluctuations in the number of particles may need to be revised for two-component systems without uniform spatial symmetry, as in the case of the neutron star crust. 

Regardless, we tried to explore the relation between these two physical observables indicated in Eq.(\ref{eq:E6}), by plotting the fractional fluctuations $\langle N\rangle^{-1}\sigma_{t}^{2}$ versus $\rho T\kappa_{T}$. Note that in an attempt to mimic the grand-canonical ensemble, all quantities in the above expression have been properly scaled to the small central sphere.

Before discussing the results, we should note that Eq.(\ref{eq:E6}) does work for an isotropic homogeneous thermodynamic syste\,\cite{34,35,36}. The inner crust of the neutron star crust, as discussed above, contains anisotropic and inhomogeneous matter. Therefore, it is expected that the slope of $L^{3}V_{sph}^{-1}\langle N\rangle^{-1}\sigma_{t}^{2}$ versus $\rho_{sph} T\kappa_{T}$ in this prominent response function should not be unity. Our results indicate a consistent trend between them, yet the slope remains less than unity. For $y_{p}=0.1$ and $0.2$, we observed that the slope increases with the simulation temperature. The reason can be assigned to the desorption of neutrons from the isolated non-symmetric clusters/pasta phases forming a relatively isotropic homogeneous system. There is a similar trend for the slope with temperature for $y_{p}=0.4$ but the slope is much smaller than unity. The observed large difference between proton-rich ($y_{p}=0.4$) and neutron-rich ($y_{p}=0.1$ and $0.2$) environments can still be attributed to the different simulated configurations. That is the proton-rich environment remains anisotropic and inhomogeneous at higher temperatures (see Fig.\;\ref{Figure_2}).

\section{CONCLUSIONS}
Motivated by the importance of understanding transport properties in neutron stars, such as the neutrino mean-free path which controls 
the neutron star cooling rate, we have investigated the rich and complex pasta structures in the inner crust by performing semi-classical MC simulation at different temperatures, densities, and proton fractions. Once the system has been thermalized, we computed the various pair-correlation functions, which were subsequently utilized as input to compute the isothermal compressibility of the system by invoking the Kirkwood-Buff theory. Our results have unveiled that in proton-poor environments, the product $\rho T \kappa_{T}$ remained almost constant over the region of temperatures and densities explored in this study. In contrast, in a proton-rich environment with $y_{p}=0.4$---where the long-range Coulomb interaction plays a preeminent role---the isothermal compressibility develops a large peak at a critical temperature $T_{c}$. This behavior was ascribed to the existence of a critical point, indicating the onset of a phase transition.

Inspired by the possible existence of a phase transition, we proposed a method to monitor particle fluctuations in the canonical ensemble by isolating a small volume at the center of the larger simulation volume. Nucleon fluctuations in the smaller volume illustrated how the thermalized configurations evolve as a function of the MC sweep. In particular and as expected, we observed strong proton localization within clusters as compared to neutrons that can exist both in clusters as well as in a dilute neutron vapor. Moreover, we noticed how the various pasta structures dissolve as the temperature of the system becomes comparable with the nuclear binding energy. We identified various processes that encapsulate the evolution of the various structures, including free neutron motion, the jiggling of bound neutrons and protons within the clusters, free neutron absorption, and bound neutron desorption. In addition, we have observed that the variance in both proton and neutron numbers approach each other as the system becomes nearly symmetric, namely, when $y_{p}\!=0.4$. The importance of the fluctuations in the particle number is its direct impact on the static structure factor and consequently, on the neutrino mean-free path; see Eq.(\ref{eq:E6}). Indeed, the larger the variance in the particle number the larger the neutrino-pasta scattering cross-section. This may dramatically affect the neutrino mean-free path\,\cite{8} which may allow for a significant energy transfer to the nuclear medium and impact the stellar cooling rate. 

Finally, we explored the correlation between the mean-square fluctuations in the particle number and isothermal compressibility for different proton fractions. Such a study revealed that the mean-square fluctuation is proportional to the isothermal compressibility, that is, $\kappa_{T}\propto \sigma_{t}^{2}$. However, the failure to observe such a perfect correlation was attributed to the relatively small system size. In the future, we plan to work on systems containing a much larger number of particles to better monitor particle fluctuations over a wide set of thermodynamic conditions of relevance to the inner crustal region of the neutron star. In particular, we aim to quantify the precise relationship between these two quantities in the form of $\kappa_{T} = \sigma_{t}^{2} \times F(\rho, y_{p}, T)$, where $F(\rho, y_{p}, T)$ is a function of temperature, baryon density, and proton fraction. 
Exploring the connection between nucleon fluctuations and the isothermal compressibility under $\beta$-equilibrium could offer a better picture of the underlying dynamics in the inner crust of neutron stars. This aspect, however, is reserved for our forthcoming investigations. In our concluding remarks, based on the plot of dimensionless nucleon fluctuation variance against isothermal compressibility as the response function, we observe that the slope of this relationship is not unity. This implies that the inner crust of neutron stars is characterized by inhomogeneous and anisotropic properties.

In summary, the present work clarifies the existence of phase transition in the inner crust of neutron stars where the emergence of pasta structures has been diagnosed by examining the behavior of the isothermal compressibility and the mean-square fluctuations in the particle number. The insights attained in this study lead to a deeper understanding of the nuclear composition in the inner crust of neutron stars.

\section*{ACKNOWLEDGMENT}
We thank Prof. Qorbani (National Taiwan University) for useful discussions, and comments, and for helping to prepare the figures. We are also thankful to Prof. Kun Yang (Florida State University) for useful discussions. We acknowledge the High-Performance Computing (HPC) system in the Department of Physics at 
the University of Tehran, where the reported Monte Carlo simulations were performed. This material is based in part upon work supported by the U.S. Department of Energy Office of Science, Office of Nuclear Physics under Award Number DE-FG02-92ER40750.


\begin{thebibliography}{0}%
\makeatletter
\providecommand \@ifxundefined [1]{%
 \@ifx{#1\undefined}
}%
\providecommand \@ifnum [1]{%
 \ifnum #1\expandafter \@firstoftwo
 \else \expandafter \@secondoftwo
 \fi
}%
\providecommand \@ifx [1]{%
 \ifx #1\expandafter \@firstoftwo
 \else \expandafter \@secondoftwo
 \fi
}%
\providecommand \natexlab [1]{#1}%
\providecommand \enquote  [1]{``#1''}%
\providecommand \bibnamefont  [1]{#1}%
\providecommand \bibfnamefont [1]{#1}%
\providecommand \citenamefont [1]{#1}%
\providecommand \href@noop [0]{\@secondoftwo}%
\providecommand \href [0]{\begingroup \@sanitize@url \@href}%
\providecommand \@href[1]{\@@startlink{#1}\@@href}%
\providecommand \@@href[1]{\endgroup#1\@@endlink}%
\providecommand \@sanitize@url [0]{\catcode `\\12\catcode `\$12\catcode
  `\&12\catcode `\#12\catcode `\^12\catcode `\_12\catcode `\%12\relax}%
\providecommand \@@startlink[1]{}%
\providecommand \@@endlink[0]{}%
\providecommand \url  [0]{\begingroup\@sanitize@url \@url }%
\providecommand \@url [1]{\endgroup\@href {#1}{\urlprefix }}%
\providecommand \urlprefix  [0]{URL }%
\providecommand \Eprint [0]{\href }%
\providecommand \doibase [0]{https://doi.org/}%
\providecommand \selectlanguage [0]{\@gobble}%
\providecommand \bibinfo  [0]{\@secondoftwo}%
\providecommand \bibfield  [0]{\@secondoftwo}%
\providecommand \translation [1]{[#1]}%
\providecommand \BibitemOpen [0]{}%
\providecommand \bibitemStop [0]{}%
\providecommand \bibitemNoStop [0]{.\EOS\space}%
\providecommand \EOS [0]{\spacefactor3000\relax}%
\providecommand \BibitemShut  [1]{\csname bibitem#1\endcsname}%
\let\auto@bib@innerbib\@empty
\end{thebibliography}%


\nocite{*}
\begin{thebibliography}
\
	\bibitem{1}
S. J. Bell Burnell, Science \textbf{304}, 489 (2004).
	\bibitem{2}
G. Baym, H. A. Bethe, and C. J. Pethick, Nucl. Phys. A \textbf{175}, 225 (1971).
	\bibitem{3}
	M. E. Caplan and C. J. Horowitz, Rev. Mod. Phys. \textbf{89}, 041002 (2017).
	\bibitem{4}
N. Chamel and P. Haensel, Living Rev. Relativ. \textbf{11}, 10 (2008).
	\bibitem{5}
W. G. Newton, Nat. Phys. \textbf{9}, 396 (2013).
	\bibitem{6}
J. M. Lattimer and M. Prakash, Phys. Rep. \textbf{621}, 127 (2016).
	\bibitem{7}
M. Thiel, C. Sfienti, J. Piekarewicz, C. J. Horowitz, and M. Vanderhaeghen, J Phys G Nucl Part Phys \textbf{46}, 093003 (2019).
	\bibitem{8}
C. J. Horowitz, M. A. Pérez-García, and J. Piekarewicz, Phys. Rev. C \textbf{69}, 045804 (2004).
	\bibitem{9}
A. Schneider, D. Berry, C. Briggs, M. Caplan, and C. Horowitz, Phys. Rev. C \textbf{90}, 055805 (2014).
	\bibitem{10}
B. Schuetrumpf, K. Iida, J. A. Maruhn, and P. G. Reinhard, Phys. Rev. C \textbf{90}, 055802 (2014).
	\bibitem{11}
J. Piekarewicz and G. T. Sánchez, Phys. Rev. C \textbf{85}, 015807 (2012).
	\bibitem{12}
B. Schuetrumpf and W. Nazarewicz, Phys. Rev. C \textbf{92}, 045806 (2015).
	\bibitem{13}
F. J. Fattoyev, C. J. Horowitz, and B. Schuetrumpf, Phys. Rev. C \textbf{95}, 055804 (2017).
	\bibitem{14}	
M. Caplan, C. Forsman, and A. Schneider, Phys. Rev. C \textbf{103}, 055810 (2021).
	\bibitem{15}
J. A. Pons, D. Viganò, and N. Rea, Nat. Phys. \textbf{9}, 431 (2013).
	\bibitem{16}
J. Piekarewicz, F. J. Fattoyev, and C. J. Horowitz, Phys. Rev. C \textbf{90}, 015803 (2014).
	\bibitem{17}
M. E. Caplan, A. S. Schneider, and C. J. Horowitz, Phys. Rev. Lett. \textbf{121}, 132701 (2018).
	\bibitem{18}
Z. Lin, M. E. Caplan, C. J. Horowitz, and C. Lunardini, Phys. Rev. C \textbf{102}, 045801 (2020).
		\bibitem{19}
C. Horowitz, D. Berry, M. Caplan, T. Fischer, Z. Lin, W.
Newton, E. O’Connor, and L. Roberts, arXiv:1611.10226.
	\bibitem{20}
S. Burrello, F. Gulminelli, F. Aymard, M. Colonna, and A. R. Raduta, Phys. Rev. C \textbf{92}, 055804 (2015).
	\bibitem{21}
R. Nandi and S. Schramm, Astrophys. J. \textbf{852}, 135 (2018).
	\bibitem{22}
R. Shafieepour, H. R. Moshfegh, and J. Piekarewicz, Phys. Rev. C \textbf{105}, 055809 (2022).
	\bibitem{23}
J. G. Kirkwood and F. P. Buff, J. Chem. Phys. \textbf{19}, 774 (1951).
	\bibitem{24}
S. S. Avancini, L. Brito, Ph. Chomaz, D. P. Menezes, and C. Providencia, Phys. Rev. C \textbf{74}, 024317 (2006).
	\bibitem{25}
S. K. Tiwari, S. Tripathy, R. Sahoo, and N. Kakati, Eur. Phys. J. C \textbf{78}, 938 (2018).
	\bibitem{26}
A. Späh et al., Phys. Chem. Chem. Phys. \textbf{21}, 26 (2019).
	\bibitem{27}
H. E. Stanley, \textit{Introduction to Phase Transitions and Critical Phenomena} (Clarendon Press, 1971).
	\bibitem{28}
P. G. Debenedetti, F. Sciortino, and G. H. Zerze, Science \textbf{369}, 289 (2020).
	\bibitem{29}
A. Schneider, C. Horowitz, J. Hughto, and D. Berry, Phys. Rev. C 88, 065807 (2013).
	\bibitem{30}
A. Schneider, D. Berry, C. Briggs, M. Caplan, and C. Horowitz, Phys. Rev. C \textbf{90}, 055805 (2014).
	\bibitem{31}
R. K. Pathria, \textit{Statistical Mechanics}, 2nd ed. (Butterworth-Heinemann, Oxford, 1996).
	\bibitem{32}
R. Kubo, Reports on progress in physics \textbf{29}, 255 (1966).
	\bibitem{33}
B. Chakrabarty, J. Chakravarty, S. Chaudhuri, C. Jana, R. Loganayagam, and A. Sivakumar, J. High Energy Phys. 2020, \textbf{165} (2020).
	\bibitem{34}
A. R. Dulaney, S. A. Mallory, and J. F. Brady, J. Chem. Phys. \textbf{154}, 014902 (2021).
	\bibitem{35}
L. Stixrude and C. Lithgow-Bertelloni, Geophys. J. Int.	 \textbf{162}, 610 (2005).
	\bibitem{36}
R. Myhill, Geophys. J. Int.	 \textbf{231}, 230 (2022).
\end{thebibliography}
\end{document}